# Thermogravitational cycles: theoretical framework and example of an electric thermogravitational generator based on balloon inflation/deflation

K. Aouane[a,b,c,d,§], O. Sandre[b,c,*], I. J. Ford[e,~], T. P. Elson[f,°], C. Nightingale[d,#]

[a] *UPMC Université Paris 6, Licence de Physique, 4 place Jussieu 75005 Paris, France*
[b] *Univ. Bordeaux, LCPO, UMR 5629, ENSCBP 16 avenue Pey Berland, 33607 Pessac, France*
[c] *CNRS, Laboratoire de Chimie des Polymères Organiques, UMR 5629, ENSCBP 16 avenue Pey Berland, 33607 Pessac, France*
[d] *University College London, Department of Mechanical Engineering, Torrington Place, London WC1E 7JE United Kingdom*
[e] *University College London, Department of Physics and Astronomy, Gower Street, London WC1E 6BT*
[f] *University College London, Department of Chemical Engineering, Torrington Place, London WC1E 7JE United Kingdom*

§ *e-mail: kamel.aouane.13@ucl.ac.uk*

\* *e-mail: olivier.sandre@enscbp.fr*

~ *e-mail: i.ford@ucl.ac.uk*

° *e-mail: t.elson@ucl.ac.uk*

# *e-mail: c.nightingale@ucl.ac.uk*

## Abstract

Several studies have combined heat and gravitational energy exchanges to create novel heat engines. A common theoretical framework is developed here to describe thermogravitational cycles which have the same efficiencies as the Carnot, Rankine or Brayton cycles. Considering a working fluid, enclosed in a balloon, inside a column filled with a transporting fluid, the cycle is composed of four steps. Starting from the top of the column, the balloon goes down by gravity, receives heat from a hot source at the bottom, rises and delivers heat to a cold source at the top. Unlike classic power cycles which need external work to operate the compressor, thermogravitational cycles can operate as "pure power cycle" where no external work is provided to drive the cycle. To illustrate this concept, the prototype of a thermogravitational electrical generator is presented. It uses a hot source of low temperature (average temperature near 57°C) and relies on the gravitational energy exchanges of an organic fluorinated fluid inside a balloon attached to a magnetic marble producing an electromotive force of 50 mV peak to peak by the use of a linear alternator. This heat engine is well suited to be operated using renewable energy sources such as geothermal gradients or focused sunbeams.

# 1. Introduction

In recent years many new techniques to produce clean energy have been introduced, a number of which still exploit heat engines using phase-change cycles where a working fluid is compressed, heated, expanded and cooled to produce work. Compression and expansion processes are commonly achieved by the use of velocity devices, e.g. turbomachinery, or positive displacement devices, e.g. reciprocating or screw mechanisms [1]. Evaporators, condensers and pumps are also required to operate in closed cycles where the working fluid is continuously circulated and does not need replenishment. The thermal energy input from renewable sources is generally obtained from different sources such as hot aquifers in the earth crust for a geothermal generator or concentrated sunbeams for solar thermal power generation. Traditional water-steam engines use high temperature sources, since below 370°C, the thermal efficiency becomes uneconomic [2]. In order to operate at lower temperatures, and therefore access a wider range of renewable energy sources, water must be replaced by lower boiling temperature fluids such as alkanes or fluoroalkanes [3], [4]. For example, such fluids are used in organic Rankine cycles (ORC) for applications like geothermal energy conversion, biomass combustion, ocean thermal energy conversion or low grade waste heat recovery [5], [6], [7], [8] or energy storage [9].

A further direction for development that has received attention involves the combination of buoyancy with an external source for power production, for example in marine power plants where buoys use the vertical movements of ocean waves to power a linear generator [7], [10], [11]. Different schemes have been proposed in the literature where the use of gravitation has been coupled to heat sources to create heat engines. In a solar balloon, absorption of heat from sunbeams modifies the density of an air filled balloon and causes its ascent to produce work [12], [13]. In a version of a magnetic fluid generator, bubbles of a non-magnetic fluid vaporized by a heat source move across a magnetic fluid (magnetized by a static magnetic field) of higher density due to buoyancy forces and produce electricity, also *via* a linear generator [14]. Interestingly, ORCs have been combined with buoyancy forces in [15] and with gravity driven compression in [16].

An extensive literature search did not yield any evidence that such cyclic patterns combining heat and gravitational work, termed thermogravitational cycles, have been analysed together to provide a common unifying framework: The present paper aims at filling this gap. Furthermore, in conventional power cycles, some external work is needed to operate the compressor. However by using thermogravitational power cycles, it is possible to obtain a "pure power cycle" where no work is provided to the cycle, *i.e.* only work is extracted from the cycle. Thermodynamic analysis and efficiency computations of thermogravitational cycles are presented in section 2.

A proof of concept of an actual thermogravitational pure power cycle, a thermogravitational electric generator, is presented in section 3. A hot source heats up an organic fluorinated fluid inside the impermeable elastic membrane of a balloon, while buoyancy force raises the inflated balloon and moves an attached magnetic marble across a solenoid. Then the organic fluid cools down in a cold region located at the top of the column and the balloon moves down, again across the coil. This upwards and downwards motion, with similarities to the thermo-convective oscillatory motion of wax blobs in a Lava lamp, powers a linear generator in closed cycles without needing any evaporator, condenser or pump.

In section 4 the implications of this study are discussed, including issues of practical implementation, and in section 5 main conclusions are given.

## 2. Theory of thermogravitational cycles

### 2.1. Definitions

#### 2.1.1. Thermogravitational power cycle

Consider a balloon which, according to the context, could be either an elastic membrane of negligible mass containing a working fluid of mass density $\rho_{\mathrm{wf}}$ or this structure attached to a weight, let's say a magnetic marble, of mass density $\rho_{\mathrm{m}}$. This balloon is placed inside a column of height $h$ filled with a transporting fluid of density $\rho_{\mathrm{tf}}$. These elements are illustrated in Figure 1. The reference vertical axis has been orientated downward and this convention will be followed for all computations through this article.

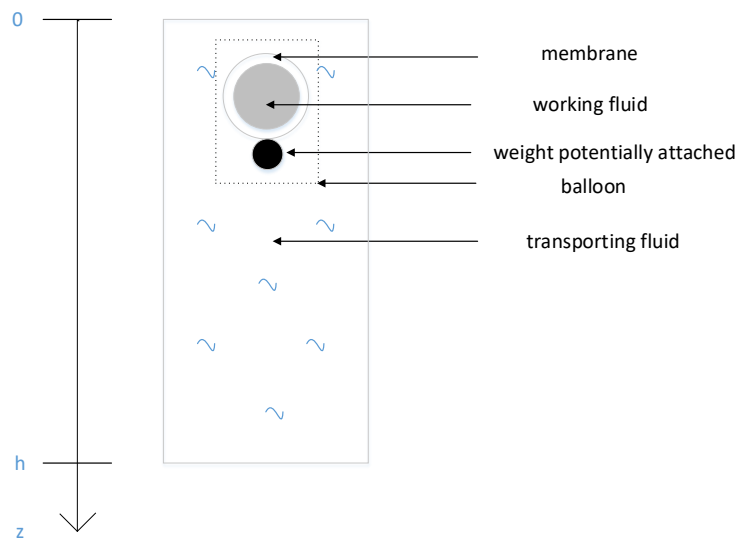

**Figure 1. Balloon, composed of a working fluid inside a membrane and a potentially attached weight (magnetic marble) inside a column of height *h* filled with a transporting fluid.**

The pressures are $P_0$ and $P_h$ at the column's top and bottom respectively. It is supposed that the pressure of the working and transporting fluids are equal for a given altitude inside the column and the balloon membrane allows no heat transfers during the fall or the rise of the balloon. A thermogravitational cycle is composed of four steps:

- 1→2: The balloon is originally at the column top. The working fluid is at the cold temperature $T_C$ and at pressure $P_0$. The balloon falls towards the bottom of the column. The working fluid experiences an adiabatic compression and reaches the pressure $P_h$ and the temperature $T_{EC}$ at the end of the compression when the balloon reaches the bottom.

- 2→3: At the column bottom, the working fluid is put in contact with the hot source at temperature $T_H$ where $T_H > T_{EC}$. The working fluid receives heat from the hot source and experiences an isobaric expansion at pressure $P_h$.

- 3→4: The balloon rises towards the column top. The working fluid experiences an adiabatic expansion and reaches the pressure $P_0$ and the temperature $T_{EE}$ at the end of the expansion when the balloon reaches the top.

- 4→1: At the column top, the working fluid is put in contact with the cold source at temperature $T_C$ where $T_C < T_{EE}$. The working fluid provides heat to the cold source and experiences an isobaric compression at pressure $P_0$.

These steps are depicted in Figure 2 beside the steps of a classic power cycle for comparison.

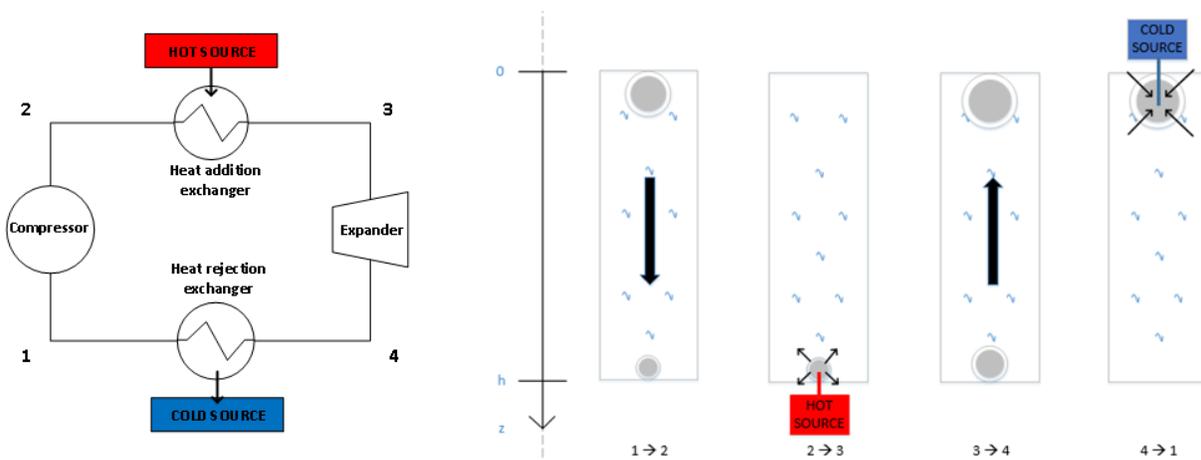

**Figure 2. Description of a classic cycle (left) and a thermogravitational cycle (right). 1->2: adiabatic compression, 2->3: hot heat transfer, 3->4: adiabatic expansion, 4->1: cold heat transfer.**
**Left figure drawn by the authors was inspired by a sketch in a popular thermodynamics textbook [17].**

The hydrostatic pressure of the transporting fluid increases with depth due to gravity. The higher the mass density of the working fluid, the lower the height of the column needed to achieve a defined compression ratio. Consequently the compression (expansion) process of the working fluid when the balloon goes down (up) can be denoted gravitational compression (expansion). It will be demonstrated that cycles that make use of gravity to drive part of the process retain the efficiencies that would normally be associated with more traditional cycles of a similar character such as the Carnot, Rankine or Brayton cycles.

### 2.1.2. Thermogravitational pure power cycle

In conventional power cycles, some external work is needed to operate the compressor. However by using thermogravitational power cycles, it is possible to obtain a "pure power cycle" where no work is input: work is only extracted from the cycle. To obtain these cycles, the density of the balloon should be higher than the density of the transporting fluid during its fall and lower than the density of the transporting fluid during its rise. If the properties of the working fluid do not allow this criterion to be satisfied, a weight could be added to the working fluid enclosed in the membrane (see Figure 1).

The mass ratio $N$ is defined as the ratio of the mass $m_\mathrm{m}$ of the weight to the mass $m_\mathrm{wf}$ of the working fluid (therefore the absence of a weight simply corresponds to the limit case $N = 0$):

$$N = \frac{m_\mathrm{m}}{m_\mathrm{wf}} \tag{1}$$

The density of the balloon, denoted $\rho_\mathrm{B}$, is then defined as the sum of the attached weight mass and working fluid mass divided by the sum of the weight volume $V_\mathrm{m}$ and working fluid volume $V_\mathrm{wf}$ :

$$\rho_\mathrm{B} = \frac{m_\mathrm{m} + m_\mathrm{wf}}{V_\mathrm{m} + V_\mathrm{wf}} = \frac{N + 1}{\dfrac{N}{\rho_\mathrm{m}} + \dfrac{1}{\rho_\mathrm{wf}}} \tag{2}$$

The balloon has a higher density than the transporting fluid if $\rho_\mathrm{B} > \rho_\mathrm{tf}$. A condition for the mass ratio $N$ is thus obtained:

$$N > \frac{\dfrac{\rho_\mathrm{tf}}{\rho_\mathrm{wf}} - 1}{1 - \dfrac{\rho_\mathrm{tf}}{\rho_\mathrm{m}}} \tag{3}$$

Similarly the balloon has a lower density than the transporting fluid if:

$$N < \frac{\dfrac{\rho_\mathrm{tf}}{\rho_\mathrm{wf}} - 1}{1 - \dfrac{\rho_\mathrm{tf}}{\rho_\mathrm{m}}} \tag{4}$$

Consequently a pure power cycle is obtained if during the fall of the balloon expression (3) is satisfied and during its rise expression (4) holds. The density of the working fluid $\rho_\mathrm{wf}$ could differ significantly during fall and rise, especially if the working fluid is in gaseous and liquid states during the rise and the fall of the balloon respectively. An experimental example of a pure power cycle is presented in section 3.

## 2.2. Side piston concept

During the rise or the fall of the balloon, its volume is modified. Assuming that the transporting fluid is incompressible, the transporting fluid surface level might be raised or lowered as a consequence. During the fall (rise), the working fluid will experience a compression (expansion) and the height of the transporting fluid will decrease (increase). These changes of height will modify the position of the centre of gravity of the transporting fluid. However to allow a simple comparison of thermogravitational cycles with other cycles, a theoretical concept, that keeps the centre of gravity of the transporting fluid at the same position during all processes, is introduced.

Consider a column with displaceable side pistons, covering both sides of the column from top to bottom. These pistons could conceivably move sideways to accommodate the variations of the balloon's volume. As depicted in Figure 3, during the fall of the balloon from 1→2:

- 1: the balloon has been inserted at the top of the column which was originally totally filled with the transporting fluid and the top pistons of the sub-system A have been displaced in order to accommodate the balloon volume.
- 1': the balloon has left the top of the column and the top pistons of the sub-system A move inwards by receiving a work input.
- 2: the balloon has reached the bottom of the column. The bottom pistons of the sub-system B are displaced outwards, delivering a work output, in order to accommodate the balloon volume.

All remaining pistons, to accommodate the balloon volume during its fall, will move out and in, resulting in a net total work equal to zero.

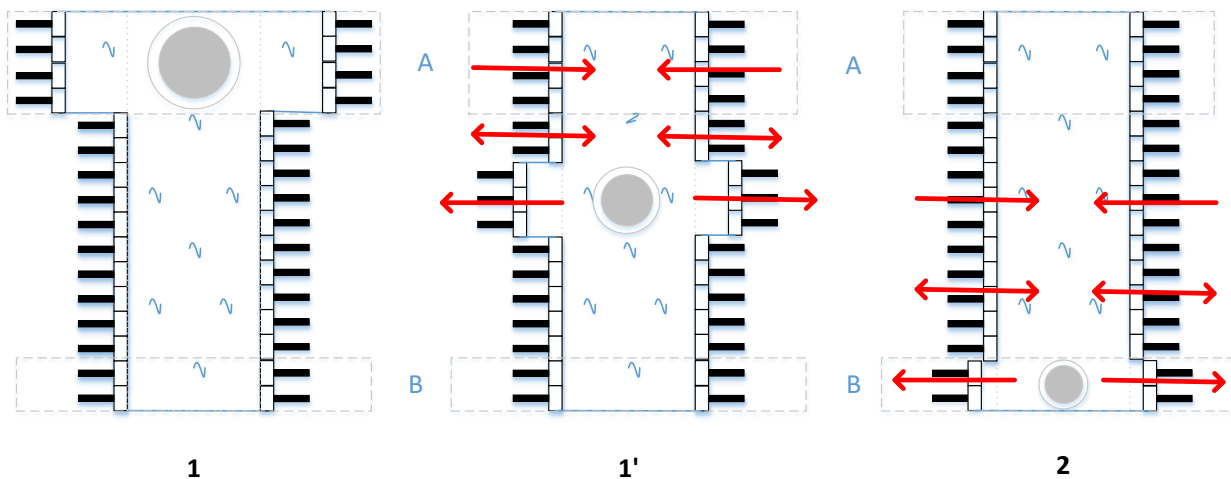

**Figure 3. Fall of the balloon and corresponding side pistons' movements. 1: start of the compression. 1': during the compression. 2: end of compression. A and B represent sub-systems at the top and bottom of the column respectively. Red arrows represent the movement of the side pistons.**

In the following, the specific work of the side pistons during the fall of the balloon is determined. Supposing sub-system A at the top of the column stays at constant pressure $P_0$, its variation of specific volume (volume per unit mass in $m^3 \cdot kg^{-1}$) during 1→2 is $\Delta v_{A,1\to2} = v_{A,2} - v_{A,1}$ where $v_{A,1}$ and $v_{A,2}$ are the specific volume of sub-system A at stage 1 and 2 respectively. The specific volume of sub-system A at each stage is the sum of the specific volume of transporting fluid and specific volume of working fluid in sub-system A. Supposing that the specific volume of transporting fluid stays the same during 1→2, and denoting $v_1$ to be the specific volume of working fluid at stage 1, it could be deduced that $\Delta v_{A,1\to2} = -v_1$. Knowing that $v_1$ is linked to the specific enthalpy $h_1$ (in $J \cdot kg^{-1}$) and specific internal energy $u_1$ (in $J \cdot kg^{-1}$) of the working fluid at stage 1 *via* the relation $v_1 = (h_1 - u_1)/P_0$, sub-system A receives the specific work $w_{A,1\to2}$ (in $J \cdot kg^{-1}$):

$$w_{A,1\to2} = \int_{A,1}^{A,2} -Pdv = -P_0 \Delta v_{A,1\to2} = h_1 - u_1 \qquad (5)$$

Repeating the above computations for sub-system B provides the specific work $w_{B,1\to2} = -(h_2 - u_2)$ due to a change of volume when the balloon arrives at the bottom of the column, where $h_2$ and $u_2$ are the specific enthalpy and internal energy of the working fluid at stage 2. Since the net resulting specific work of the remaining pistons is nil, the net work performed on the sub-systems composing the column other than sub-systems A and B is nil. Hence the net specific theoretical work $w_{1\to2,th}$ performed by all the side pistons during the compression 1→2 is finally the sum of $w_{A,1\to2}$ and $w_{B,1\to2}$. Defining $\Delta u_{1\to2}$ and $\Delta h_{1\to2}$ as the variations of specific internal energy and specific enthalpy respectively from 1→2:

$$w_{1\to2,th} = w_{A,1\to2} + w_{B,1\to2} = \Delta u_{1\to2} - \Delta h_{1\to2} \qquad (6)$$

Following similar reasoning, the net specific work of the side pistons during each of the remaining steps (heat transfers and expansion) of a thermogravitational cycle is equal to the difference between the variation of specific internal energy and the change in specific enthalpy of the working fluid as expressed in equation (6) for the compression stage. Naturally the sum of works performed by the side pistons over the cycle is equal to zero.

## 2.3. Ideal thermogravitational power cycles

### 2.3.1. Thermogravitational power cycle efficiency

In the following description, some simplifying assumptions are made. Firstly, during the rise and fall of the balloon, no heat transfers are allowed (adiabatic assumption) and the frictional losses of the balloon with respect to the transporting fluid are neglected. Secondly, the working fluid is assumed to reach the desired temperatures at the bottom and top of the column. To compute the specific gravitational work $w_{1\to2,grav}$ on the system during the compression process, the first law of thermodynamics is used (where the sign convention is the one where all energy transfers to the system are positive):

$$w_{1\rightarrow2,\text{grav}} + w_{1\rightarrow2,\text{th}} + q_{1\rightarrow2} = \Delta u_{1\rightarrow2} + \Delta e_{\text{p},1\rightarrow2} \tag{7}$$

The specific heat transfer $q_{1\rightarrow2}$ (in J·kg$^{-1}$) is equal to zero as the process is adiabatic, the variation of specific potential energy is $\Delta e_{\text{p},1\rightarrow2} = -gh$ and the specific work performed by the side pistons $w_{1\rightarrow2,\text{th}}$ is derived from equation (6). Replacing each term in equation (7), it is found that:

$$w_{1\rightarrow2,\text{grav}} = \Delta h_{1\rightarrow2} - gh \tag{8}$$

Following a similar reasoning, the specific gravitational work during the expansion process is $w_{3\rightarrow4,grav} = \Delta h_{3\rightarrow4} + gh$. Regarding the specific heat transfers with the hot and cold sources, they occur at constant pressure and are then equal to the variation of the working fluid specific enthalpy after and before the heat transfers, *i.e.*:

$$q_{2\rightarrow3} = \Delta h_{2\rightarrow3} \ \text{ and } \ q_{4\rightarrow1} = \Delta h_{4\rightarrow1} \tag{9}$$

The efficiency of a power cycle is defined as the ratio between the algebraic sum of the cycle work (both gravitational and from the side pistons) and the heat provided by the hot source:

$$\eta = \frac{-\sum_{\text{cycle}}(w_{\text{th}} + w_{\text{grav}})}{q_{2\rightarrow3}} \tag{10}$$

Inserting the corresponding values, the efficiency of a thermogravitational power cycle is found:

$$\eta = \frac{(h_3 - h_4) - (h_2 - h_1)}{h_3 - h_2} \tag{11}$$

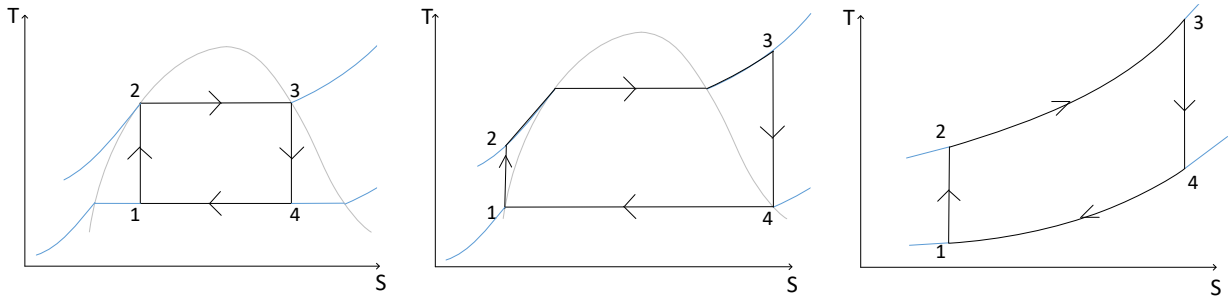

**Figure 4. Carnot cycle (left), Rankine cycle (middle) and Brayton cycle (right). 1->2: isentropic compression, 2->3: isobaric heat transfer with hot source, 3->4: isentropic expansion, 4->1: isobaric heat transfer with cold source. Grey and blue curves describe conditions of phase coexistence and isotherms respectively.**

Each of the steps (compression, expansion and heat exchanges) can occur while the working fluid is in different thermodynamic states: gas, liquid or a mixture of coexisting gas and liquid. Depending on the different processes the working fluid goes through, thermogravitational cycles can be classified as thermogravitational Carnot, thermogravitational Rankine or thermogravitational Brayton cycles which, as will be demonstrated in the two following sections, have the same efficiencies as their classic counterparts namely the Carnot, Rankine and Brayton cycles. As a reminder, these cycles are presented in a temperature entropy diagram in Figure 4.

### 2.3.2. Thermogravitational phase-change cycles

If the case where the heat exchanges occur at constant pressure and constant temperature is considered, and by defining $s$ as the specific entropy of the working fluid, some simplifications can be made: $h_3 - h_2 = q_{2\rightarrow3} = T_H(s_3 - s_2)$ and $h_1 - h_4 = q_{4\rightarrow1} = T_C(s_1 - s_4)$. Since processes 1→2 and 3→4 are adiabatic, equation (11) could be simplified as:

$$\eta = 1 - \frac{T_C}{T_H} \qquad\qquad (12)$$

This is the expression of the Carnot cycle efficiency and consequently such a thermogravitational cycle is referred to as the thermogravitational Carnot cycle.

If the compression process occurs in the liquid region of the phase diagram, the efficiency found in equation (11) is the Rankine cycle efficiency. This thermogravitational cycle is consequently referred to as the thermogravitational Rankine cycle. To illustrate this cycle, several simulations have been realized (see Figure 5) with water as a transporting fluid and three different fluoroalkanes as working fluids: perfluoropentane $C_5F_{12}$ (boiling temperature at 1 bar of 29°C), perfluorohexane $C_6F_{14}$ (boiling temperature at 1 bar of 56°C) and perfluoroheptane $C_7F_{16}$ (boiling temperature at 1 bar of 83°C). Only pure power cycles are considered and since these fluids in the liquid state have higher densities than water, no weight is needed to be attached to the membrane enclosing the working fluid to allow the balloon to fall. The efficiencies following equation (11) have been computed for temperatures and pressures up to 150°C and 10 bar respectively. The temperatures at the end of the compression and expansion of the working fluid have been found by considering isentropic processes. At the top of Figure 5, for a given height, and at the bottom of Figure 5, for a given temperature, the minimum possible boiling temperature and the maximum possible height have been chosen respectively to keep the working fluid at the gas and liquid state during the rise and the fall of the balloon. Consequently it could be noticed at the top of Figure 5 that the efficiency curve of $C_7F_{16}$ stops at around 40 meters since for higher columns temperatures above 150°C are needed to vaporize the fluid.

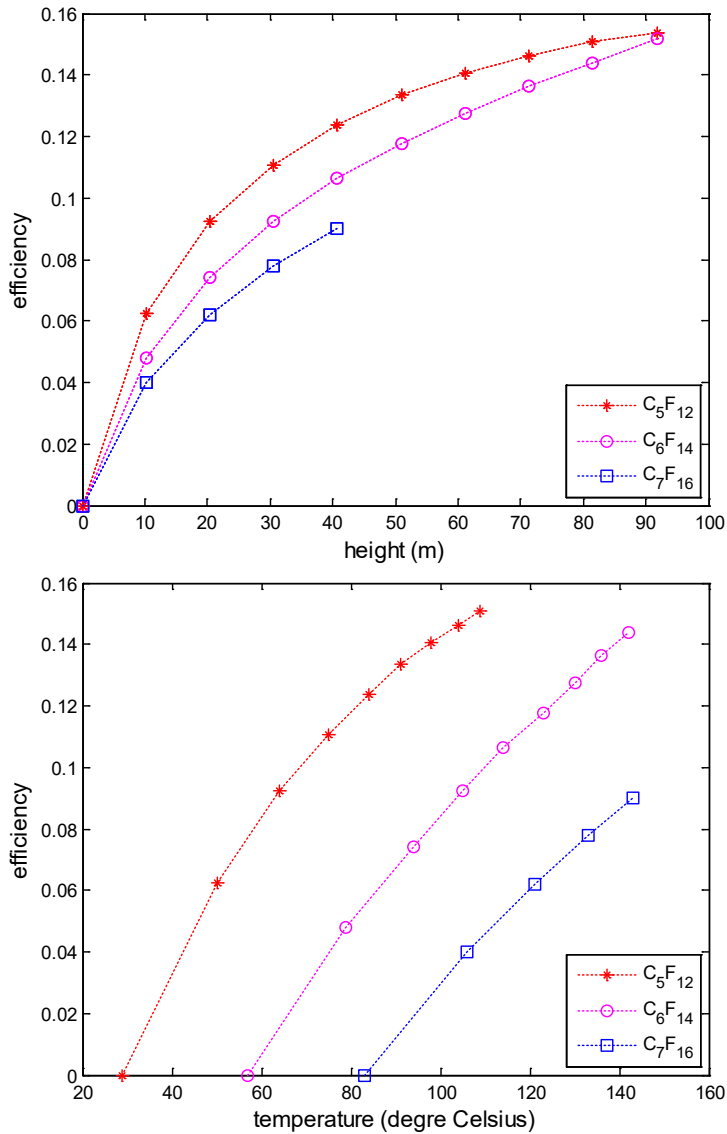

<span style="color:blue">**Figure 5. Efficiency VS height (top) and efficiency VS temperature (bottom). Simulations have been carried out for three different working fluids ($C_5F_{12}$, $C_6F_{14}$ and $C_7F_{16}$) following equation (11) for a thermogravitational Rankine cycle with hot source temperatures and pressures up to 150°C and 10 bar respectively. The cold source temperature is fixed at 20°C. The working fluid is kept in the gas and liquid state during the rise and the fall of the balloon respectively.**</span>

CHEMCAD (version 6.5.6) was used to calculate vapour fraction, density, enthalpy and entropy data for its library components perfluoropentane, perfluorohexane, and perfluoroheptane over a series of temperatures and pressures required for the calculations above. CHEMCAD [18] is a chemical process simulator which uses the DIPPR thermodynamic and physical property database [19] for its library components. Methods used by CHEMCAD to calculate thermodynamic properties are described by Edwards in [20]. Enthalpies are calculated from the ideal gas heats of formation, integrating the specific heats and incorporating heats of vaporisation where appropriate, and entropies are calculated in the usual manner from $\Delta S = \int \frac{\Delta Q}{T} \, \mathrm{d}T$. Vapour density is obtained from the ideal gas law modified to include the compressibility factor. CHEMCAD uses DIPPR correlation parameters for temperature dependent properties where available. These are available for perfluoropentane for liquid density, vapour pressure, liquid heat capacity, heat of vaporization and ideal gas heat capacity. DIPPR parameters are not available for perfluorohexane and perfluoroheptane and so CHEMCAD's standard methods are used for these two components, including liquid density (API), vapour pressure (Antoine equation), and ideal gas heat capacity (polynomial).

### 2.3.3. Thermogravitational gas cycle

Finally we consider the case where all processes occur while the working fluid remains as a gas (assumed to be perfect), *i.e.* the working fluid does not go through any change of phase. By using the relationship $\Delta h = c_p \Delta T$ and introducing the pressure ratio $r_p$ for adiabatic processes 1→2 and 3→4:

$$r_p = \frac{P_2}{P_1} = \left(\frac{T_2}{T_1}\right)^{\frac{\gamma}{\gamma-1}} \quad \text{and} \quad r_p = \frac{P_4}{P_3} = \left(\frac{T_4}{T_3}\right)^{\frac{\gamma}{\gamma-1}} \tag{13}$$

By converting equation (11) from enthalpy changes to temperature changes and substituting equation (13), the efficiency can be written as:

$$\eta = 1 - \frac{1}{r_p^{\frac{\gamma-1}{\gamma}}} \tag{14}$$

This is the efficiency of the Brayton Cycle and consequently this thermogravitational cycle is referred to as the thermogravitational Brayton cycle. Given a hot and cold source at temperature $T_H$ and $T_C$ respectively, there exists an optimal pressure ratio $r_{p,opt}$ to maximise the specific net work output. Differentiating the specific net work output with respect to $r_p$ and equating the result to zero, it can be shown that [21]:

$$r_{p,opt} = \left(\frac{T_H}{T_C}\right)^{\frac{\gamma}{2(\gamma-1)}} \tag{15}$$

If a liquid is used as a transporting fluid, pure power cycles for the thermogravitational Brayton cycles cannot generally occur without a mass attached to the working fluid enclosed in a membrane to allow the latter to fall during the compression stage. Supposing that the working fluid is a perfect gas with an adiabatic exponent $\gamma$, the conditions on the mass ratio $N$ of expressions (3) and (4) could be written as:

$$\frac{\frac{R T_H \rho_{tf}}{P_h M_{wf}} - 1}{1 - \frac{\rho_{tf}}{\rho_m}} > N > \frac{\frac{R T_C \rho_{tf}}{P_0 M_{wf}} - 1}{1 - \frac{\rho_{tf}}{\rho_m}} \tag{16}$$

If the transporting fluid is a liquid, $P_{\mathrm{h}} = P_0 + \rho_{\mathrm{tf}} gh$. Replacing $P_{\mathrm{h}}$ in equation (16) leads to an upper limit on the height $h$ of the column that will allow the cycle to occur by only extracting work, *i.e.* $h < \frac{P_0}{\rho_{\mathrm{tf}} g}\left(\frac{T_{\mathrm{H}}}{T_{\mathrm{C}}} - 1\right)$. Using expression (14) an upper efficiency is found for a pure thermogravitational Brayton cycle:

$$\eta < 1 - \left(\frac{T_{\mathrm{C}}}{T_{\mathrm{H}}}\right)^{\frac{\gamma-1}{\gamma}} \tag{17}$$

To illustrate this point, Figure 6 presents the optimal Brayton efficiency using the optimal pressure ratio of equation (15) and the upper efficiency for a pure power Brayton cycle using expression (17) as a function of the hot source temperature. The working fluid is air considered as a perfect gas with $\gamma = 1.4$ and the cold source temperature has been chosen to be equal to 20°C.

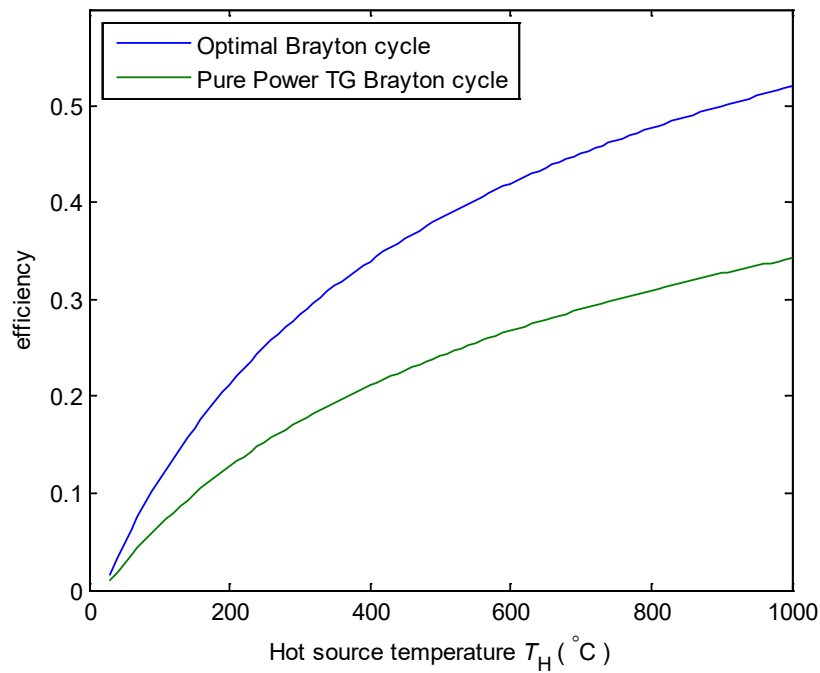

**Figure 6. Efficiency comparison of an optimal and a pure power thermogravitational (TG) Brayton cycle depending on the hot source temperature $T_{\mathrm{H}}$. A pure power cycle requires no work input for the compression process. The working fluid is air and the cold source temperature is equal to 20°C.**

From the curves depicted in Figure 6, it could be inferred that for a given hot source temperature one could obtain a better cycle efficiency by providing some external work rather than using a pure power thermogravitational Brayton cycle. However this is not mandatory as will be evidenced in following part.

# 3. Thermogravitational electric generator

## 3.1. Experiment

The experimental set-up of a thermogravitational electric generator, which employs a thermogravitational Rankine cycle, is presented in Figure 7. A working fluid volume of perfluorohexane ($C_6F_{14}$) is introduced through the needle of a syringe into a nitrile elastomer bag cut from a glove finger (denoted "2"), where air has been completely removed with a membrane pump before sealing by a tight knot. This volatile fluoroalkane liquid is marketed by 3M under the brand name Fluorinert™ FC-72® (also provided by Sigma-Aldrich under reference 281042, 99% grade) with a liquid density $\rho_{wf} = 1680 \text{ kg} \cdot m^{-3}$, a molar mass $M_{wf} = 0.338 \text{ kg} \cdot mol^{-1}$ and a boiling point $T_b \approx 56°C$ at normal temperature and pressure conditions [22]. The bag is attached to a strong spherical FeBNd magnet of radius $R_m = 6.5 \text{ mm}$, density $\rho_m = 7500 \text{ kg} \cdot m^{-3}$ and magnetization $M \approx 850 \times 10^3 \text{ A} \cdot m^{-1}$ (labelled "1") according to the specifications given by the provider, Superaimants-Europe, France.

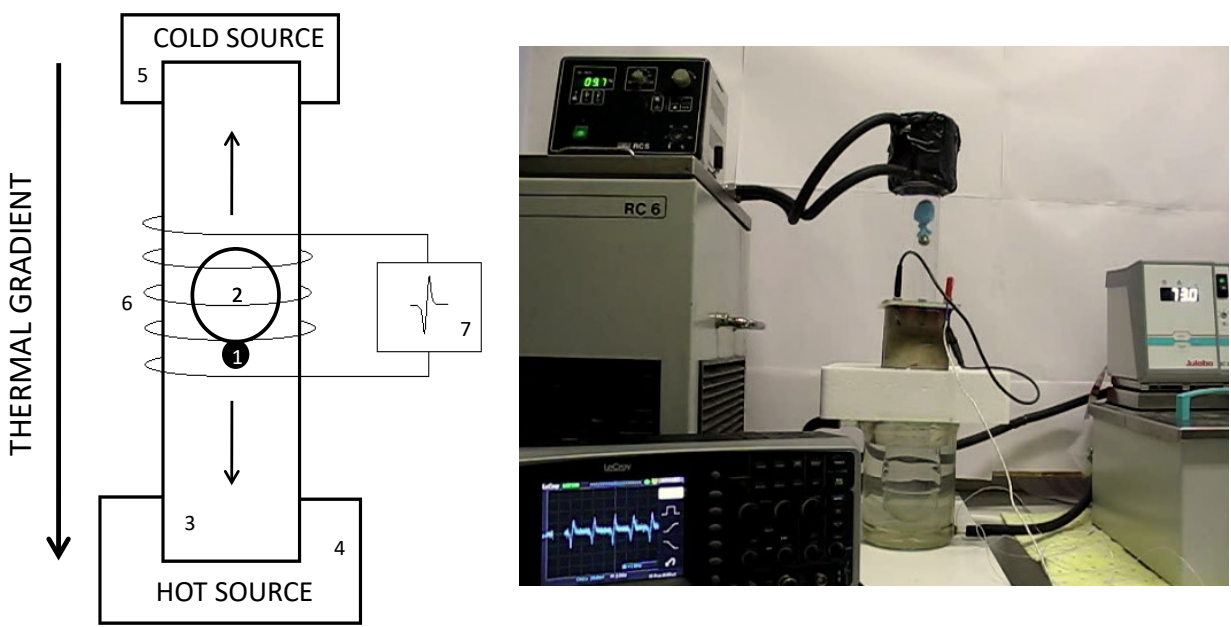

**Figure 7.** Scheme and picture of the experimental set-up (1: magnetic marble, 2: elastic bag, 3: water-filled column, 4: hot source, 5: cold source, 6: solenoid, 7: oscilloscope). A video of the setup in action is provided in supplementary data.

The balloon {bag + magnet} is introduced into a cylindrical column of radius $R_c = 4.10^{-2} \text{ m}$ and height $h = 0.48 \text{ m}$, containing water as a transporting fluid (labelled "3") of density $\rho_{tf} \approx 1000 \text{ kg} \cdot m^{-3}$ and viscosity $\eta_{tf} \approx 5 \times 10^{-4} \text{ Pa} \cdot s$ (near 50°C). The column is surrounded at the bottom by a hot water-jacket at 73°C (labelled "4") and at the top by a cold one at 10°C (labelled "5"). Due to these hot and cold water circuits, a thermal gradient is achieved inside the column between a measured hot temperature $T_H \approx 57°C$ at the bottom and a cold temperature $T_C \approx 51°C$ at the top. This temperature difference is lower than between the two water-jackets due to convection flows within the water column reminiscent of the Rayleigh-Bénard instability. When heated at the bottom near to its boiling point temperature, the perfluorohexane droplet vaporizes and inflates the bag, decreasing the balloon density. Once the latter falls below the density of water, the balloon rises according to Archimedes' principle. During its progression towards the top, the vaporized fluorocarbon gas condenses as the surrounding water gets colder. The bag progressively deflates and once the balloon density becomes larger than water density, it falls down. The cycle repeats again and again, resulting in oscillations of the magnet between the top and the bottom of the column, as indicated by the arrows in Figure 7. Each time the magnet goes through the coil (labelled "6"), the latter experiences a variation in the magnetic flux: according to the Faraday-Lenz law, an electromotive force (*e.m.f.*) is induced and detected by an oscilloscope (labelled "7").

A pure power cycle has been designed for this proof of concept. On one hand, expression (3) is always satisfied since the density of the working fluid in the liquid state is higher than the water density during the fall of the balloon. On the other hand, during the rise of the balloon, following expression (4), a maximum mass ratio $N = 87$ has been computed corresponding to a minimum working fluid volume in the liquid state of 59 μL . Therefore a volume $V_{wf} = 80$ μL of perfluorohexane in the liquid state was used to ensure that the balloon reaches the cold zone at the top of the column. With this volume of working fluid enclosed in the membrane, as depicted on Figure 7, the inflated bag has a radius $R_b \approx 1.2$ cm (measured on the digital images by comparison with the diameter of the magnetic marble).

This experiment has also been performed with decafluoropentane as a working fluid ($C_5H_2F_{10}$, boiling point range $53 - 55°C$, 60% purity grade provided by DuPont™ under the brand name Vertrel® XF and available from Sigma-Aldrich under reference 94884) where similar oscillations of the balloon in the column were observed. However thermodynamic tables were not available for this fluid that is a mixture of two stereoisomers (89% erythro / 11% threo) with very different boiling points [23] and therefore we chose instead pure perfluorinated alkanes as working fluids to compare with theory.

## 3.2.  Electrical and mechanical analysis

### 3.2.1.  Electromotive force calculation

As illustrated in Figure 8 using a spherical coordinate system with unit vectors ($\boldsymbol{u_r}, \boldsymbol{u_\theta}, \boldsymbol{u_\varphi}$), the magnetic vector potential at a point $K$ of an electrical conductive loop (L) of radius $R_l$ and altitude $z_l$, created by a homogeneous spherical magnet (m) assimilated as a magnetic dipole of volume $V_m$, magnetization $M$ and magnetic moment $\boldsymbol{m} = MV_m\boldsymbol{u_z}$, falling or rising along the $z$ axis with unit vector $\boldsymbol{u_z}$, is [24]:

$$\boldsymbol{A}(K) = \frac{\mu_0}{4\pi} \frac{\boldsymbol{m} \wedge \boldsymbol{u_r}}{r^2} \qquad (18)$$

Still referencing to Figure 8, $\boldsymbol{m} = m(\cos\theta\,\boldsymbol{u_r} - \sin\theta\,\boldsymbol{u_\theta})$, $z_r = z_l - z_m$, $r = \sqrt{R_l^2 + z_r^2}$ and $\sin\theta = R_l/r$. The expression above can be rewritten as:

$$\boldsymbol{A}(K) = \frac{\mu_0}{4\pi} \frac{mR_l}{(R_l^2 + z_r^2)^{3/2}} \boldsymbol{u_\varphi} \qquad (19)$$

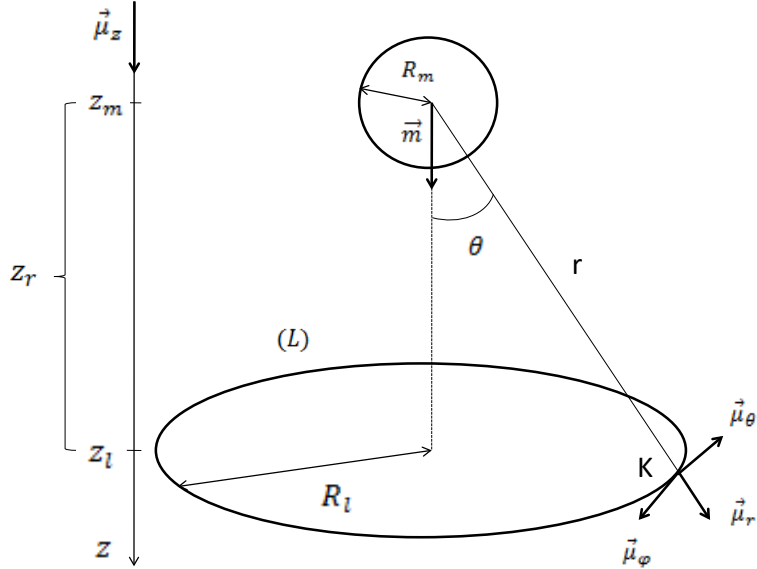

**Figure 8. Geometry for the electromotive force *(e.m.f.)* calculation showing the falling marble (m) above the loop (L).**

By definition the electromotive field is $\boldsymbol{E_m} = \partial \boldsymbol{A}/\partial t$ [24], and keeping in mind that the magnet altitude $z_m$ is dependent on time, one obtains (the dot denotes the time derivative):

$$\boldsymbol{E_m}(K) = \frac{3\mu_0}{4\pi} \frac{mR_l z_r \dot{z}_m}{(R_l^2 + z_r^2)^{5/2}} \boldsymbol{u_\varphi} \tag{20}$$

The electromotive force (*e.m.f.*) created in a section of a conductor delimited by two points A and B is $e = \int_A^B \boldsymbol{E_m}.\, \boldsymbol{dl}$ [24]. To obtain the induced electromotive force created by the magnet over the loop (L), we use $\boldsymbol{dl} = R_l \mathrm{d}\varphi \boldsymbol{u_\varphi}$ where $\varphi$ is the azimuthal angle in the spherical coordinate system. Hence the electromotive field can be integrated between 0 and $2\pi$ to give the voltage:

$$e = \frac{2\pi\mu_0 MR_l^2 R_m^3 z_r \dot{z}_m}{(R_l^2 + z_r^2)^{5/2}} \tag{21}$$

Calculations by other methods in the literature gave similar results [14],[25], apart from a $4\pi$ pre-factor instead of $2\pi$ in [14] that is most likely erroneous.

### 3.2.2. Magnetic marble speed and position

Equation (21) requires the time dependent speed of the magnet (equal to the balloon speed) $\dot{z}_m$ and the altitude $z_m$. The balloon is modelled for hydrodynamic drag as a perfect sphere of radius $R_B$ and cross-sectional area $A = \pi R_B^2$. This sphere experiences hydrodynamic forces which can be modelled for the relevant regime of the Reynolds number comparing convective to viscous forces. The "pipe flow" model is well suited to the situation, where $\bar{\dot{z}}_m$ denotes the average speed of the magnet [26]:

$$Re = \frac{\bar{\dot{z}}_m R_c \rho_{tf}}{\eta_{tf}} \tag{22}$$

For a Reynolds number between $10^3$ and $10^5$, the system is in the Newtonian regime and a quadratic drag force $\boldsymbol{F_d}(\dot{z}_m) = \frac{1}{2} C_d \rho_{tf} A \dot{\boldsymbol{z}}_{\boldsymbol{m}}^{\boldsymbol{2}}$ should be considered, where the drag coefficient for a sphere is constant in this Reynolds number interval: $C_d \approx 0.4$ [26]. The sphere is subject to gravitational and drag forces. By applying Newton's second law, expressions for the magnet speed $\dot{z}_m$ and altitude $z_m$ are obtained. By considering the case when the sphere starts from an initial speed equal to zero and by introducing the reduced quantities $\tilde{g} = g(1 - \rho_{tf}/\rho_B)$ and $\alpha = \sqrt{-0.15 \, \rho_{tf}/(\rho_B \, R_B \tilde{g})}$, the analytical solution has been found by Timmerman and van der Weele [27]:

$$\dot{z}_m(t) = \frac{1}{\alpha} \tanh(\alpha \tilde{g} t) \tag{23}$$

$$z_m(t) = h + \frac{1}{\alpha^2 \tilde{g}} \ln[\cosh(\alpha \tilde{g} t)] \tag{24}$$

### 3.3. Results

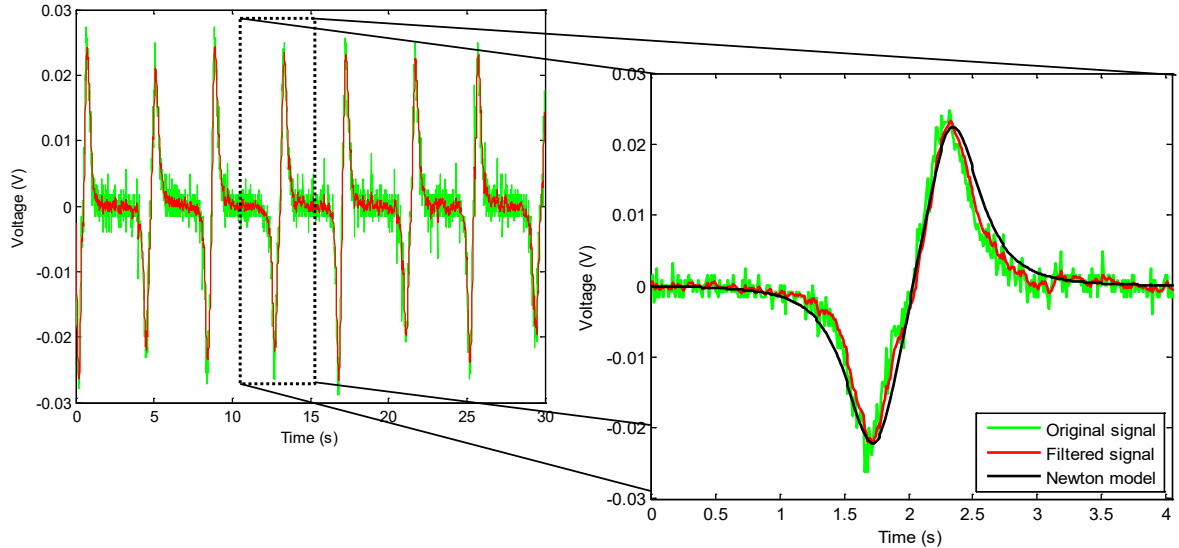

Figure 9. Signal acquisition on the oscilloscope. The magnification shows the *e.m.f.* produced during a single balloon rise. The green and red lines represent respectively the experimental signal before and after electronic filtering with a low band pass *RC* filter of cut-off frequency 5Hz ($R$=10 k$\Omega$, $C$= 3.3μF). The black line is the analytical curve also called "Newton model" in the text, in which the only adjustable parameter is the balloon radius $R_B = 1.22$ cm (best fit value that is very close to the balloon radius of 1.2 cm measured on images).

The signal observed on the oscilloscope during an acquisition time of 30 seconds is plotted in Figure 9. The oscillation period is defined as the time needed by the balloon to return to a given altitude of the column, passing by the top and the bottom of the column. Therefore the oscillation period is measured by the duration necessary for two *e.m.f.* pulses to appear in Figure 9, or equivalently by the average time (very regular) measured between two "clicks" heard when the marble touches the bottom of the glass column in the movie of the oscillations provided as supplementary data. The period averaged over different measurements under the same conditions is $\bar{T} \approx 8.3$ s (for the thermal gradient and volume of working fluid in the balloon provided in the experimental part). In Figure 9, the magnified view of the *e.m.f.* corresponds to the signal during one rise of the balloon, starting from the bottom and proceeding to the top of the column. The rise took $\bar{T}/2 \approx 4.15$ sec, resulting in an average magnetic marble speed of $\bar{z}_m = 0.12$ m · s⁻¹. The Reynolds number calculated using equation (22) is $Re \approx 10^4$ which justifies the use of the Newton regime valid for $Re$ between $10^3$ and $10^5$. To obtain the theoretical curve of the *e.m.f.* against time, called the "Newton model" in Figure 9, the set of equations was solved numerically for a discrete time variable range from 0 to 4.15 s. At each time, the values of the magnet speed calculated analytically by equation (23) and of the magnet altitude from equation (24) were inserted in equation (21) to find the *e.m.f.* induced in one loop. Then the electromotive forces of the loops located at different radii $R_l$ and altitudes $z_l$ were added to obtain the total *e.m.f.* induced in the coil. This one is defined by 1044 turns of conducting wire of 0.8 mm cross-section diameter, spaced by 0.06 mm of insulating layer, in a total height of 75 mm. Each of the 87 levels of wiring contains 12 loops of increasing radius $R_l$ from 38 mm (internal radius) to 48 mm (external radius).

For a periodic signal, the electrical energy produced during one period is calculated as $\bar{E} = \frac{1}{R} \int_0^T e^2(t)\mathrm{d}t$ where $R = 9.6\ \Omega$ is the measured coil resistance. Since random noise would cause a slight overestimate of this calculation, an electronic *RC* low pass filter with a cut-off frequency of 5Hz was applied to the signal. Using a numerical integration of the filtered signal depicted on Figure 9, and averaging the result over several periods, an average energy $\bar{E} = 58\ \mu J$ was obtained, corresponding to an average power $\bar{P} = \bar{E}/\bar{T} = 7\ \mu W$.

By comparison, the energy needed to vaporize the working liquid is $Q = m_{wf}L_v$ where $L_v = 88\ J \cdot g^{-1}$ is the latent heat of vaporization of perfluorohexane [22], giving $Q = 12$ J. The experimental efficiency of the system $\eta_{exp} = \bar{E}/Q$ is $4.8 \times 10^{-6}$. In this estimate, the energy needed to operate the two thermal sources was not taken into account, since in foreseen applications this energy is expected to be found externally from industrial waste heat, solar or geothermal gradient heat sources for example. From a theoretical point of view, computing the efficiency following equation (11) with $T_H$ and $T_C$ of the experiment and the corresponding enthalpy data for the perfluorohexane, the value of $2.6 \times 10^{-5}$ is found which is around five times higher than the experimental value. This difference is easily explained as the experiment differed from the theoretical model in several ways: The linear alternator did not harvest all the available energy in the system over the column height, the elastic membrane certainly allowed some heat transfers during compression and expansion, and in addition some frictional losses occurred during the rise and fall of the balloon.

The oscillations of the balloon can last several hours (at least up to 5 h). However, after some time, the balloon progressively remains in an inflated state at the top of the column. We hypothesise that this phenomenon is ascribed to a composition change of the fluorocarbon. If some plasticizers of the elastomer migrate into the working fluid and react with it, the molecular composition can change. The boiling temperature can decrease, explaining why some of the liquid no longer condenses in the cold region. In potential applications, this drawback should be circumvented by a careful choice of the materials (fluid and elastic bag) to eliminate interactions between them even after numerous thermal cycles. However, the prototype described here proved the concept of pure thermogravitational cycles.

## 4. Discussion

The thermogravitational cycles described here are power cycles that consume heat and produce a net work output. However, these cycles could be reversed to work as a heat pump or a refrigerator to respectively provide heat or maintain coolness by consuming work. The four steps described for the thermogravitational cycles remain the same. However at the bottom of the column during step 2→3, the working fluid donates heat to the hot source and experiences an isobaric compression. At the top of the column during step 4→1, the working fluid receives heat from the cold source and experiences an isobaric expansion. As for the power cycles, considering the different processes the working fluid goes through, the thermogravitational reversed cycles can be classified as thermogravitational reversed Carnot, thermogravitational reversed Rankine or thermogravitational reversed Brayton cycles, amongst others. Thermogravitational reversed cycles are described in more detail in the M.Sc. thesis of one of the authors [29].

Besides the thermodynamic properties of the working fluid [3], [4], legislative measures and regulations should be considered too: due to potential danger of long chain fluoroalkanes for the environment [29], [30], [31] a volatile perfluoroether, perfluoropoly(ether) or fluoro-olefin (unsaturated) could be used instead.

To increase the energy output of the thermogravitational electric generator, the parameters of equation (21) giving the *e.m.f.* should be optimized, for example by using a bigger magnet that better fills the interior of the solenoid, *i.e.* minimizing the conducting loop-to-magnet radii ratio $R_l/R_m$. One should also keep in mind Lenz's principle causing a braking force induced by the magnetic field created by the eddy currents in the coil, which opposes the penetration of the magnetic marble through it. The surrounding medium could be changed by using a transport fluid of larger mass density than water to increase the Archimedes' force and to achieve a higher magnet speed during the rising step, while maintaining a low viscosity to maintain a high Reynolds number: some non-volatile fluorinated oils with mass densities around 2 g·cm$^{-3}$ might be good candidates, apart from their rather high cost.

To obtain high efficiency thermogravitational power and reversed cycles, more development has to be carried out on the heat exchanges at the top and bottom of the column. Different technologies for gravitational compression / expansion have been proposed in the literature [10], [12], [15] but their efficiencies need to be better assessed. The selection process for the working and transporting fluids taking into account the heat sources, space requirement, desired output and efficiency has to be identified. Last but not least, cost analysis of the global system with their different components has to be quantified and compared to the economics of existing technologies.

# 5. Conclusion

A framework for the use of gravitational forces and heat exchanges to produce work has been proposed. The cycles have been referred to as thermogravitational cycles and, if producing a net work output, they are a form of power cycles. These cycles are reversible and consequently could consume work to heat or cool a closed space, as in a heat pump or a refrigerator. Taking into account the work of the conceptual side pistons, introduced in order that the gravitational energy of the transporting fluid is held constant, it has been demonstrated that, following the nature of the different processes the working fluid goes through, the thermogravitational cycle efficiencies are the same as well-established thermodynamic cycles namely the Carnot, Rankine or Brayton cycles or their reversed counterparts. In Table 1, the main disadvantages and advantages of thermogravitational concepts are summarized.

| Disadvantages | Advantages |
|---|---|
| System is not compact: high columns | Possibility of wet compression /expansion to approach the Carnot efficiency |
| Slow gravitational compression and expansion:<br>- low power outputs<br>- adiabatic condition is more difficult to approach | Can operate even under very low hot source temperature, according to the specifications of the organic fluid used |
| Achieving efficient heat exchanges at the top and bottom of the column could be challenging | Possibility to have pure power cycles |

Table 1. Disadvantages and advantages of thermogravitational cycle concepts.

The theory has been illustrated with the presentation of a thermogravitational electric generator prototype. Such a system could potentially be used with any gradient of temperature, for example with geothermic heat, solar energy, or industrial waste heat. The underlying physical features of the system were modelled quantitatively; in particular the resulting *e.m.f.* was predicted analytically, without any adjustable parameter. Taking into account limitations such as the neglect of friction or the partial harvesting of energy due to the coil dimensions, the efficiency of the experiment was reasonably close (by five times lower) to the predicted theoretical efficiency.

The theory described in this paper and the presentation of the thermogravitational electric generator provide the basis for a better understanding of thermogravitational cycles thereby acting as a starting point for extension of these concepts to renewable energy conversion or storage.

## Acknowledgements

The authors thank Nabil Mongalgi for his help with the signal acquisition and the electronic filtering and Dr A. Thabti for bringing reference [9] to their attention.